# AI and Democracy's Digital Identity Crisis

Shrey Jain,
*Harvard Getting Plurality Lab,*
Boston, United States,
shreyjaineth@gmail.com

Connor Spelliscy,
*Decentralization Research Center*
Toronto, Canada
connor@statlerlabs.com

Samuel Vance-Law,
*Decentralization Research Center*
Berlin, Germany
samuelvancelaw@gmail.com

Scott Moore
*Decentralization Research Center*
Toronto, Canada
scott@publicworks.fm

*Abstract*—AI-enabled tools have become sophisticated enough to allow a small number of individuals to run disinformation campaigns of an unprecedented scale. Privacy-preserving identity attestations can drastically reduce instances of impersonation and make disinformation easy to identify and potentially hinder. By understanding how identity attestations are positioned across the spectrum of decentralization, we can gain a better understanding of the costs and benefits of various attestations. In this paper, we discuss attestation types, including governmental, biometric, federated, and web of trust-based, and include examples such as e-Estonia, China's social credit system, Worldcoin, OAuth, X (formerly Twitter), Gitcoin Passport, and EAS. We believe that the most resilient systems create an identity that evolves and is connected to a network of similarly evolving identities that verify one another. In this type of system, each entity contributes its respective credibility to the attestation process, creating a larger, more comprehensive set of attestations. We believe these systems could be the best approach to authenticating identity and protecting against some of the threats to democracy that AI can pose in the hands of malicious actors. However, governments will likely attempt to mitigate these risks by implementing centralized identity authentication systems; these centralized systems could themselves pose risks to the democratic processes they are built to defend. We therefore recommend that policymakers support the development of standards-setting organizations for identity, provide legal clarity for builders of decentralized tooling, and fund research critical to effective identity authentication systems.

*Keywords—AI, digital identity, decentralization, democracy, policy*

## I.  Introduction

When you want to communicate with your member of Congress, you don't walk down to their office and have a conversation; you send them an email, tweet at them, or sign an online petition. Modern democracies rely upon the integrity of these sorts of digital communications between people and their governments. However, AI systems have become sophisticated enough to enable malicious actors to severely undermine the integrity of those communications through the impersonation of citizens at scale.

AI-enabled tools like large language models (LLMs) allow a small number of individuals to run effective disinformation campaigns of an unprecedented scale by taking advantage of outdated and vulnerable identity authentication systems. Such attacks could quickly have a material impact on the democratic process, and governments would likely react by attempting to employ identity systems that distinguish between citizens and non-citizens (or bots) online. Unfortunately, when employing such systems, governments tend to rely on centralized solutions, which could themselves pose significant risks to the stability of democracies.

In this paper, we explore some of the known threats that AI presents to our existing digital identity systems in the democratic context and examine a potentially critical second-order effect of those threats: governments creating centralized identity authentication systems. We provide details on a representative set of different types of identity attestations, discuss their pros and cons, and propose several practical policy recommendations.

## II.  AI Puts Democratic Systems at Risk

Modern AI models are increasingly capable of mimicking targeted individuals [1], [2], [3], [4], [5]. Users of these models can influence policymakers by impersonating constituents and can influence populations via compelling disinformation.

Models have already been used to impersonate constituents at such large scales as to convince policymakers that they represent consensus views. In 2020, researchers Sarah Kreps and Douglas L. Kriner sent over 30,000 emails to more than 7,000 state legislators—50 percent written by AI and 50 percent written by students. These emails claimed to be from constituents, but legislators and their staff were unable to differentiate between the AI-generated and human-generated correspondence, with AI content eliciting only 2 percent fewer responses [6]. Since then, AI capacities have improved dramatically [7], increasing the viability of this threat. If such tactics were adopted more broadly, the fundamental communication channels between constituents and their representatives could break down, jeopardizing the integrity of representative systems.

AI-generated disinformation in the media is also creating significant challenges for policymakers and officials. In March 2023, AI-generated photos appearing to show the arrest of former President Trump went viral, causing outrage [8], and, in April, celebritiesdeaths.com, one of dozens of sites written almost entirely by AI, pronounced the death of President Biden [9]. In May, an AI-generated image of an alleged attack on the Pentagon briefly caused the stock market to dip [10], and, in June, Governor Ron DeSantis's campaign used AI to generate images in an attack ad against former President Trump [11]. Whether in product reviews, content farms, or news sites, the use of AI to generate photos and text is growing rapidly [12]. Senate Majority Leader Chuck Schumer recently told an audience at the IBM Headquarters that one of his biggest concerns was that AI deep fakes would make "people lose faith in democracy altogether" [13].

### III. IDENTITY ATTESTATIONS

Impersonation and disinformation campaigns are not uniquely enabled by AI [14], but the rapid development of such systems and the exponential scaling of associated threats require novel solutions. As it becomes impossible to differentiate between human and AI-generated content, we need more than ever to be able to confirm not simply *what* is being said but also *who* is saying it.

To identify that a statement is being made by a particular individual online, as opposed to a bot, we rely on the identity attestations tied to the source of the statement. Identity attestations are verifiable claims from a trusted issuer to a receiver. A simple example of a digital identity attestation is the gray checkmark on X (formerly Twitter), which is issued by X to accounts X believes represent government/multilateral organizations or officials.

With AI systems able to mimic human interaction, resilient, privacy-preserving identity attestations can drastically reduce instances of impersonation and make disinformation easier to identify and potentially hinder. These systems, even those specifically designed to detect bots, are under constant pressure from developments in AI: GPT-4 has succesfully persuaded a human to solve a CAPTCHA [15], though it is already the case that bots tend to be able to solve CAPTCHA tests more quickly than humans [16]. The development of AI-resistant identity attestations is therefore vital in differentiating between human and AI-generated content and in verifying sources of information. In the following section, we discuss different forms of digital identity attestations that governments might employ to distinguish between citizens and bots online.

Identity attestations are often categorized as either centralized or decentralized. For instance, a driver's license issued by the State of California might be considered a centralized attestation, while a social club's acceptance of a new member based on the collective endorsements of established members might be considered a decentralized attestation. However, while this binary delineation of centralized and decentralized attestations can be useful, it fails to categorize attestations accurately, as they generally exist on a spectrum between centralization and decentralization. Some digital identity attestations are more centralized in certain dimensions (e.g., architecturally, politically, and logically) than others, and there are situational advantages and disadvantages to different forms of attestation across this spectrum [17]. While government-issued identity can be efficient, for instance, it lacks the robustness and resilience of an attestation that is predicated on independent recommendations from multiple high-trust individuals or groups.

There is no universally accepted bright-line test for centralization in the context of identity attestations, but there are at least two key dimensions to consider: the technological composition of the attestation, and how it is governed. By setting criteria around both dimensions, we can establish a more robust measurement of centralization across various identity attestations and set better standards to assess their resilience in the context of an AI-driven world.

Technological decentralization concerns the resilience of the core infrastructure underpinning an identity attestation. An evaluation of the technological decentralization of a system might include assessing the number and diversity of nodes in a system, replacement cost of system components, number and diversity of contributors to the codebase, and whether the codebase uses proprietary or open standards. The goal of these criteria is to determine how much the system itself *depends* on any given component or person.

Governance decentralization refers to the management of the attestations related to a given system. Attestations that tend towards centralization are controlled by a smaller number of parties, whereas attestations that tend towards decentralization distribute control amongst a larger and uncorrelated group [18]. These criteria may help define the organizational structure governing how an attestation system runs, or what kinds of decisions lead to a valid attestation.

Identity attestations that tend towards decentralization across both technological and governance dimensions can allow users control over their own identity data (including the ability to selectively disclose information) and can limit manipulation by dispersing various forms of control among a large but manageable collusion-resistant group [19]. They are also harder to attack because they lack sensitive central points and are less likely to fail because they rely on multiple components [17].

However, the current technological complexity of these attestations can make them hard to use, and a lack of standardization can lead to significant friction, especially as the number of different potentially incompatible attestations increases. The lack of a central authority can also make it difficult to resolve disputes or recover lost credentials. There are ongoing efforts to improve and standardize decentralized identity attestations, such as work on Verifiable Credentials (VCs) and Decentralized Identifiers (DIDs) [20].

In evaluating identity attestations, it's essential to analyze both their technological and governance layers. These layers provide a foundation upon which we can determine the degree of centralization or decentralization inherent in an attestation. With this understanding, we can discern two primary strategies to achieve decentralization, either in full or in part:

*1) Individual Control*: This involves an individual attestation being governed by a mechanism that is inherently decentralized. Such attestations stand alone in their decentralized nature and don't rely on external factors or combinations with other attestations.

*2) Collective Reliance*: This strategy emphasizes the power of aggregation. By having a multitude of attestations and leveraging advanced tools for their aggregation, we can create a decentralized framework. In such a setup, no single attestation becomes indispensable. If one attestation is absent or deemed unreliable, its void can be filled by relying on others in the set. By understanding and applying these strategies, stakeholders can better navigate the spectrum of centralization and decentralization, optimizing for both security and user agency.

## IV. THE CURRENT SPECTRUM OF DIGITAL IDENTITY ATTESTATIONS

Below is an overview of a selection of digital identity attestations that, we believe, provide a representative sample of the spectrum of attestations currently in use. This is by no means an exhaustive survey of different examples of attestations in a particular category, and there is also overlap between categories. By understanding how each of these identity attestations is positioned across the spectrum of decentralization, we can gain a better understanding of the costs and benefits of these categories. In some instances, we have given multiple examples of attestations in a particular category to provide more context.

### A. Government Identities (e-Estonia and China's Social Credit System)

Government digital identities refer to official, state-backed digital attestations or credentials given to citizens, which allow them to authenticate themselves online for various public services, transactions, and activities. These systems aim to streamline service delivery, enhance security, and foster trust in digital interactions between a government and its citizens. However, they also introduce meaningful risks related to the trustworthiness of the government in question.

e-Estonia is the Estonian government's state-backed attestation, and it links many of the state's key functions through a single online platform, including voting, education, justice, policing, and taxes [21]. While these credentials are usually considered highly centralized, Estonia's implementation achieves a modicum of technological decentralization. The government's data platform links encrypted servers, leveraging blockchain-based Keyless Signature Infrastructure, and stores information locally, making the platform less susceptible to a complete data breach [21].

In terms of governance, e-Estonia is fundamentally a permissioned system controlled by the Estonian government, and, as such, is strictly centralized. As such, the system has limited transparency and accountability, and there are no technical impediments to the government using the system as it chooses, even when such use might be illegal.

China's social credit system is a centralized digital attestation in development that assigns points to citizens based on behavior. These metrics are designed to adjudicate financial, judicial, commercial, and societal trustworthiness, and can be affected by actions ranging from credit scores to jaywalking [22]. Citizens with high scores can expect priority for education and employment and short wait times in hospitals, while citizens with low scores may have reduced access to public services and face public shaming [23]. Some version of this system is currently employed across 80 percent of China's provinces and its corporate iteration assesses 33 million businesses [23].

### B. Biometric Identities (Worldcoin)

Biometric identity attestations rely on physical markers like fingerprints, iris scans, and voice patterns. These kinds of markers are extremely difficult to forge and there are a wide range of options for streamlined verification. The use of biometric data, however, has its own set of risks: biometric identifiers rarely meet accessibility standards and risk excluding a range of vulnerable populations. Worse, if biometric data is compromised it cannot be changed like traditional passwords, making centralized databases of this information highly vulnerable to attack and the consequences of these attacks potentially life-altering for users.

Worldcoin is a digital identity and financial services application that leverages iris scans to prove unique personhood. In exchange for providing their data, users receive a cryptocurrency by the same name [24]. Worldcoin was co-founded by Sam Altman and was specifically built in anticipation of a future in which digital identity was at risk as a result of the increased capacity of AI to mimic humans online and the need for universal basic income (UBI) given employment disruption [25].

Worldcoin aspires to provide global biometric identities in a more decentralized way than other historical biometric systems in order to try to avoid well-studied risks and trade-offs for users within these systems [26]. Technologically, it attempts to achieve this by using the Ethereum blockchain and open-sourcing parts of its software stack. However, not all of Worldcoin's tooling is open source, including the hardware device used for iris scans and the retrieved iris data itself which sits on centralized servers [27].

From a governance perspective, Worldcoin is controlled by its initial team, developers, and foundation [28]. Although Worldcoin expects decentralized governance to play a "key role" [28], its attempts to do this have had questionable results. Firstly, 25 percent of all of Worldcoin's tokens have been allocated to the core team and investors [29], enabling a small group to greatly impact decisions. Secondly, the dynamics associated with earning Worldcoins have already incentivized malicious interactions between those responsible for administering scans and those receiving them. Orb operators in multiple regions have been reprimanded for attempting to fake scans, stealing tokens from those they were scanning, and even (rarely) attempting to scan unwilling participants [30]. As a result, countries like Kenya have at least temporarily banned Worldcoin while they further investigate its operations [31].

Although biometric identity comes with strong guarantees, its implementations either through centralized companies like CLEAR or more decentralized attempts like Worldcoin still introduce significant risks that need to be addressed. While these approaches may eventually play an important role in efficiently solving for online identity, they must be carefully designed and deployed given the significant risks that the use of such personal data presents.

### C. Federated Identities (OAuth)

Federated identity systems are not themselves attestations but rather aggregators of sets of attestations from multiple authorized applications or services. Rather than requiring a new username and password for each service, federated systems provide a framework in which users can choose from a pre-selected group of trusted platforms and organizations that will attest to their personhood on their behalf. This approach simplifies the complexity of the set of attestations required and improves overall user experience by reducing password fatigue.

OAuth (Open Authorization) is the most popular open authentication system for federated identity and allows users to choose from a wide range of providers to access third-party applications. By "signing in" with their Google or X (formerly Twitter) account, users can bootstrap their identity without providing any new information about themselves.

From a technological perspective, federated systems like OAuth use centralized entities to attest to a user's identity, making management of the federation tractable, but these attestations are only as good as the organizations they include. Given the consolidation and increasing lack of public trust in Internet platforms over the past decade [32], as well as the increasing research that shows a failure of these organizations to protect against Sybil attacks [33], there is reason to believe these solutions are not a panacea and require extensive further research [34].

From a governance perspective, OAuth is managed by working groups within the Internet Engineering Task Force (see RFC 6749 and RFC 6750) and has a relatively strong process in place for managing existing stakeholders. However, given the system design risks above, this process may need to be updated to handle a larger number of organizational requests to become attestors, and to help improve the inclusivity of the review process.

The decentralization of federated identity systems is emerging. As an example, projects like Gitcoin Passport [35] and SpruceID [36] have already begun engaging with these working groups [37]. In addition, initiatives like "sign-in with Ethereum" [38] that leverage pseudonymous identifiers like those provided by the Ethereum Name Service [39] are in progress and starting to breathe new life into these standards.

### D. Web of Trust-Based Identities (EAS)

Projects like EAS (and SpruceID mentioned above) are exploring a range of related decentralized attestation architectures, with a focus on fully peer-to-peer vouching via webs of trust [40]. In EAS, users create and revoke attestations relating to specific activities or traits of an individual allowing for more specific permissions to be granted on the Ethereum blockchain. These approaches are based on the longstanding idea that trust can be measured as a weighted graph, where a person is representative of the sum of their positive attestations from others (their reputation) similar to nodes in a graph [41].

With a large enough data set, these attestations can become highly robust ways of measuring personhood, if not foundational to the notion of identity itself [42]. However, similar to large-scale social networks these models are vulnerable to Sybil attacks and as such any attestation must be able to be proven or *revoked* based on clear, measurable outcomes.

Many web of trust projects are governance minimized, open source, and forkable. In the case of EAS, since the network of attestations is on Ethereum, anyone can access and use the graph without needing to continue using the original code. More research on zero-knowledge proofs is necessary to better ensure the privacy of this data.

### V. IDENTITY AUTHENTICATION SYSTEMS

An identity authentication system is the set of attestations required to access a given resource. For example, at a bar, a bartender will ask for a single attestation—your driver's license—to allow access. A bank, however, will likely require multiple attestations, such as a username and password, an authentication app, and a code sent to your email. In each case, a provider requires a set of attestations to allow access, which the user may or may not be able or willing to provide. This relationship between user, attestations, and provider makes up an identity authentication system.

The design and the implementation of an identity authentication system are shaped by the community using that system, which determines the number and nature of attestations necessary. Weyl et al. [43] described this type of dynamic relationship as "Intersectional Social Data (ISD)," an identity that evolves and is connected to a network of similarly evolving identities that verify one another. In this type of system, each entity contributes its respective credibility to the attestation process, creating a larger, more comprehensive set of attestations, theoretically enhancing the strengths and mitigating the weaknesses of any single attestation type. Critically, the robustness and privacy of ISD hinge on having multiple elements of an individual's identity tied to one another without allowing would-be attackers the ability to connect those elements or trace them to a user. An ideal system of this type could theoretically allow a provider to perfectly assess the trustworthiness of a user without that user having to share any personal data with a provider.

A perfect system of this type does not yet exist for digital identity at scale, but some identity aggregators are beginning to realize the potential of ISD. Gitcoin Passport, for example, is an aggregator that enables communities to require multiple centralized or decentralized attestations to authenticate user identity through a single tool. As the number of attestations a user has grows, and as the reputation of individual identity attestors increases, Gitcoin can be more confident that participants across their ecosystems are unique humans. Currently, these attestations must be renewed every 90 days, creating a dynamic system that can adapt over time and continually reinforce trust. This aggregator is particularly relevant because it was created to deal with threats that bots pose to democratic processes, similar to the AI threats outlined above. Gitcoin, a Decentralized Autonomous Organization (DAO) that runs public good grants funding using a mechanism that requires Gitcoin to be able to distinguish between bots and humans online, created the Gitcoin Passport to prevent Sybil attacks—in this case, attacks by bots posing as distinct humans. Other DAOs have created similar tooling in response to attacks on the democratic processes vital to their key functions, whether establishing the rules of the organization, recording and executing decisions made by members, or managing community treasuries [44].

Identity authentication systems built on the premise of ISD provide perhaps the best approach to authenticating identity and protecting against some of the threats to democracy that AI can pose in the hands of malicious actors. It can make impersonation by bots difficult to impossible and can reliably be used to adjudicate whether information comes from trustworthy sources.

### VI. DANGERS OF GOVERNMENTS' CENTRALIZED APPROACH TO IDENTITY

We have outlined some of the threats posed by AI impersonation and disinformation above. However, certain responses to those dangers represent their own risks to democracies.

If the attacks we have highlighted above significantly impact democratic processes (and we believe they will), governments will likely employ centralized digital authentication systems based on comprehensive personal and biometric data; however, we caution against the use of these systems given how powerful they can become and the risks they may pose to the stability of a democracy.

To start with, valuable personal data stored in a centralized platform increases the likelihood of a hack—an enormous risk given the value of the extensive personal data stored in such a system. However, a centralized digital identity system presents a potentially more fundamental risk to democracy. Such a system provides existing and future governments with a powerful new tool to stifle democratic rights and target dissenters. Below we survey some of the digital identity authentication systems currently perpetuating or vulnerable to such abuses.

The most extreme example of a centralized government identity system may be China's social credit system (introduced above), which aims to monitor and maintain extensive data on the activities of every person in China with the stated goal of increasing "trust" within Chinese society [22]. The data gathered, through a broad range of sources, is sometimes accessible to a range of regulators through a centralized database or sometimes accessible only to individual regional or central government authorities [23]. Advances in technology, like facial recognition software and AI, have increased the data-gathering capacity and robustness of the system, particularly in light of China's installation of 200 million surveillance cameras [23].

While the credit system is still evolving and has yet to be fully implemented, the ambition is for each Chinese citizen to be assigned a social credit score based on the government's ongoing assessment of their activities. Some of the activities believed to result in a lower social credit ranking include bad driving, smoking in non-smoking zones, posting fake news, and frivolous purchases [45]. However, they also include posting anti-government content and participating in "illegal" protests [22]. These social credit scores act as attestations that can be used by government and non-government stakeholders to take actions against citizens with low social credit scores, including travel bans, school bans, reduced employment opportunities, and public shaming, or to reward citizens with high social credit scores [23]. Significantly, the system uses "artificial intelligence to provide early warning of risky actors in need of extra regulatory attention" and has led to 23 million people being blacklisted from plane or train travel [23].

It may seem that this type of authoritarian system would never be legal and could never be implemented in democratic countries in North America or the EU, but many of these countries have started initiatives that have already comprehensively centralized the personal data of their citizens. In so doing, they have created tools that have limited or no technological barriers to be repurposed to perform functions similar to the social credit system. For instance, the e-Estonia platform collects identity data including name, date of birth, fingerprint, address, citizenship, and personal codes [46]. Similar systems have been implemented in Denmark and Finland [47], [48].

While e-Estonia and related platforms seem to emphasize privacy and transparency, government affiliates using e-Estonia, including police officers, can illegally access information about users stored on the platform [49]. e-Estonia employs a permissioned blockchain that allows citizens to see who has accessed their information; however, that permissioned blockchain is controlled by, and could be unilaterally altered by, the Estonian government [50]. If Estonians elected a government that did not respect the legal protections of its citizens, that government could use the platform for a variety of nefarious purposes without recourse, including targeting opposition or punishing women retroactively for receiving abortions. Furthermore, if the Estonian government were to be overthrown by a foreign aggressor, like Russia, or taken over through a coup, that aggressor could gain access to the e-Estonia system, use it to cement its power, and remove all privacy protections.

The Taliban's takeover of identity authentication systems in Afghanistan is a cautionary tale about the dangers of centralized databases for identity attestations. After invading Afghanistan, foreign governments built out comprehensive stores of biometric and personal data of Afghans for various purposes. When the U.S. military pulled out of Afghanistan, the Taliban gained access to some of these systems, which held information including iris scans, fingerprints, photographs, occupations, home addresses, and names of relatives [51], [52]. One such system, funded by the U.S. government, contained records of at least 2.5 million people in Afghanistan [51]. There is evidence that the Taliban has used this biometric information to target its perceived opponents, including those who collaborated with foreign governments during the occupation [51].

VII. RECOMMENDATIONS FOR POLICYMAKERS

Given the urgency and potential scope of the risks identified, it is critical for policymakers to fast-track and support research on identity attestations and authentication systems. We fear that if AI tooling enables large-scale disruptions of democratic processes, policymakers will make reactionary decisions to adopt centralized identity authentication tooling without anticipating the risk of building such systems. We encourage governments to start researching these risks now so that they can be ready with effective solutions in the case of such a disruption.

Below is a set of policy recommendations that could apply to most jurisdictions, though we include some discussion specific to the United States.

*A. Support the Development of Standards-Setting Organizations for Identity*

Technological standards-setting organizations, like the Internet Engineering Task Force (IETF) and the World Wide Web Consortium (W3C) continue to help set standards for digital identity to improve a variety of its features, including but not limited to its security, reliability, and compatibility. There can be an overlap between these types of organizations, and industry may or may not play a significant role in their creation and management. Still, governments should play a role in the creation and maintenance of related organizations, and have done so successfully with other standards organizations in the past, including with the IETF, which was funded exclusively by U.S. government grants and meeting fees between 1986 and 1997 [53].

Creating initiatives to revive and incentivize working groups around topics like digital identity that include a diverse set of stakeholders is critical to ensure we develop effective identity attestations and authentication systems. The U.S. government can support these initiatives by funding or signaling support for existing standard-setting organizations like those above, in addition to emerging groups like the Decentralized Identity Foundation and the Plurality Research Group, which have already added to the conversation around standards-setting but do not currently have the level of influence necessary to bring these standards to industry and the public. The United States could also help by explicitly acting as a convener of these organizations as well as other industry stakeholders.

*B. Provide Legal clarity for Builders of Decentralized Tooling*

Builders of decentralized identity attestations, particularly blockchain-based ones, often face legal uncertainty that limits innovation. For instance, tooling like Gitcoin Passport and SpruceID as described above have the potential to effectively aggregate identity attestations at scale, and with the advent of zero-knowledge proofs, these decentralized solutions can be strongly privacy-preserving without losing their resilience. However, a lack of legal clarity limits the advancement of blockchain technology, and those building and using the technology in the United States are afraid of reprisal from regulators [44]. As an example, the CFTC has recently advanced the position that builders working for a DAO, which have been leaders in iterating on decentralized identity systems within their own organizations, should face unlimited liability for the actions taken by that DAO [54]. Policymakers and regulators need to provide the builders of blockchain technology with the confidence to experiment and build without fear of reprisal and liability.

We recommend U.S. policymakers pass legislation that provides legal clarity and fosters innovation among good actors building decentralized identity systems using blockchain technology. The European Union has already started to do this in the form of their recently passed Markets in Crypto Asset (MiCA) legislation, but North America is lagging behind. One example of U.S. legislation that shows promise is the Financial Innovation and Technology for the 21st Century Act [55]. While the bill is in relatively early stages and subject to change, it addresses some of the critical issues for blockchain builders, including clarifying the role of regulators, identifying when a digital asset is a security or a commodity, and addressing how blockchain companies can raise funding.

*C. Fund Research Critical to Effective Identity Authentication Systems*

There is limited research on the efficacy of different types of digital identity authentication systems and the long-term implications for governments that employ them. As stated above, we see clear dangers and pitfalls in the reliance on centralized attestations and systems that governments currently prefer. However, there has not yet been enough research to identify further weaknesses or assess the limitations or dangers of large-scale decentralized systems.

A pivotal distinction in identity authentication is the difference between "proof of human" and "proof of membership" protocols. The former ensures that every participant in a system is a unique human, aiming to prevent automated systems or malicious actors from artificially inflating numbers or skewing results. The latter, however, verifies the authenticity of a participant's membership in a particular community or group. This distinction becomes crucial in specific democratic contexts. For instance, in Taiwan's case, merely ensuring a "proof of human" wouldn't defend against threats to their democratic system. If 20 million people from the People's Republic of China, potentially influenced by the CCP, were to validate their humanity, they could flood Taiwan's digital commons, skewing public discourse. What's essential in such a scenario is "proof of membership" to ensure that participants are genuine Taiwanese citizens, safeguarding the process from external influence. Given these nuances, we suggest that governments fund research focused on user studies to understand how these systems work in different contexts before they are deployed at scale. As more data becomes available, ongoing assessment of these systems and their practical effects is crucial.

Given the pressing public sector need, some of the research funding allocated by the Biden Administration to organizations like the National Science Foundation for artificial intelligence research should be distributed to researchers focused on different types of identity authentication tooling. The private sector is already funding related research and building tooling in response to AI's impact on digital identity outside of the scope of democratic systems. However, this work may not be adequately tailored to the government's needs or be completed with necessary haste given existing threats.

VIII. CONCLUSION

AI-enabled tools will quickly have a material impact on the verifiability of identities in some key democratic processes which is of immediate concern given the upcoming US federal election. Robust digital identity authentication systems would help us differentiate between humans and AI, and efforts to improve and integrate these systems into our interactions with the digital sphere will help protect our democratic systems from AI-generated harm. However, if governments employ centralized digital identity systems, those systems could themselves have long-term destabilizing effects on the democracies they were built to protect. Before developing and deploying digital identity authentication systems, governments should fund and support initiatives to identify different forms of identity authentication systems and study their benefits and risks.

ACKNOWLEDGMENTS

We would like to thank our reviewers: Bruce Kogut, Blair Levin, Steven Nam, Morva Rohani, Helena Rong, Derek Slater,

Statements and views expressed in this publication are solely those of the authors and do not imply endorsement by the reviewers and their respective organizations